\documentclass[12pt]{iopart}

\usepackage{graphicx}
\begin{document}

\title[Probing Local Structure in Glass by the Application of Shear]{Probing Local Structure in Glass by the Application of Shear}

\author{Nicholas B. Weingartner and Zohar Nussinov}

\address{Department of Physics, Washington University in St. Louis, One Brookings Drive, St. Louis, MO 63130-4889, USA}

\ead{weingartner.n.b@wustl.edu}
\ead{zohar@physics.wustl.edu}
\vspace{10pt}
\begin{indented}
\item[]January 2016
\end{indented}

\begin{abstract}
The glass transition remains one of the great unsolved mysteries of contemporary condensed matter physics. When crystallization is bypassed by rapid cooling, a supercooled liquid, retaining amorphous particle arrangment, results. The physical phenomenology of supercooled liquids is as vast as it is interesting. Most significant, the viscosity of the supercooled liquid displays an incredible increase over a narrow temperature range. Eventually, the supercooled liquid ceases to flow, becomes a glass, and gains rigidity and solid-like behaviors. Understanding what underpins the monumental growth of viscosity, and how rigidity results without long range order is a long-sought goal. Many theories of the glassy slowdown require the growth of static lengthscale related to structure with lowering of the temperature. To that end, we have proposed a new, natural lengthscale- "the shear penetration depth". This lengthscale quantifies the structural connectivity of the supercooled liquid. The shear penetration depth is defined as the distance up to which a shear perturbation applied to the boundary propagates into the liquid. We provide numerical data, based on the simulations of $NiZr_2$, illustrating that this length scale exhibits dramatic growth and eventual divergence upon approach to the glass transition. We further discuss this in relation to percolating structural connectivity and a new theory of the glass transition.
\end{abstract}

%
%
%
%
%

\section{The Glass Transition and the Shear Penetration Depth}
\subsection{Solids, Liquids, and Supercooled Liquids}
Liquids flow, solids do not. This is the fundamental distinction between the two states of matter. If a force is applied to a liquid, its constituents will rearrange until a new equilibrium state is reached. By contrast, a solid will deform elastically under small stresses, but large scale atomic rearrangements will not occur. The fundamental physical variable that quantifies this difference is the viscosity, a measure of a material's resistance to flow. The viscosity of a liquid is finite, wheareas the viscosity of a perfect crystal is infinite. In futher contrast with a liquid, a crystalline solid is characterized by time-independent elastic and shear moduli. The elastic forces that resist deformation in response to applied stresses and thermal fluctuations in a crystalline solid, are a consequence of the long range structural order. This structural order is acquired at the uniquely defined melting point where there is a thermodynamic driving force for structural change: the free energy of the system decreases through macroscopic atomic ordering. 

Just above the melting temperature, as the liquid is coooled, its viscosity increases and is well approximated by an Arrhenius law,
\begin{equation}
\label{Arrhenius}
\centering
\eta(T)=\eta_0 e^{\frac{E}{k_B T}} ,
\end{equation}
where $\eta_0$ is the extrapolated value of viscosity at infinite temperature, $E$ is a nearly constant function of T (playing the role of activation energy of atomic rearrangement), and $k_B$ is the usual Boltzmann constant. Eventually, when the liquid is cooled to its melting temperature, $T_m$, a first order phase transition occurs, and the liquid becomes a crystalline solid. This process requires a finite amount of time, the nucleation time, however, and can therefore be thwarted. If a liquid is cooled sufficiently quickly, crystallization can be bypassed and the liquid enters a metastable state known as the supercooled liquid. As a supercooled liquid is further cooled, its viscosity begins to increase in an incredible fashion, by up to 16 decades over a temperature window as small as a few hundred Kelvin. A "glass transition" temperature, labeled $T_g$, is eventually reached at which point the viscosity is so great (~$10^{13}$ Poise) that any large scale atomic rearragnement (flow) ceases on physically meaningful timescales (order of ~ 100s). On daily time-scales, beneath $T_g$, the supercooled liquid responds rigidly to external perturbations and is deemed a glass. If the viscosity of the supercooled liquid continued to increase in an Arrhenius fashion down to $T_g$, the glass transition would not have been so mysterious. In such a case, there would exist a constant energy barrier to local rearrangement that would become exponentially harder to cross as kinetic energy is removed from the supercooled liquid on cooling. The supercooled liquid wouldn't flow on experimentally meaningful timescales and the apparent rigidity would be entirely kinetic in origin. However, for all supercooled liquids, a departure from Arrhenius dynamics is observed. For all liquids, there is an apparent temperature dependence to the activation energy \cite{7,8}, and the degree of departure from an Arrhenius viscosity is quantified in the fragility parameter of Angell \cite{7}. Despite a wide and rich phenomenology that cannot be enumerated here \cite{1,2,3,4,5,6}, understanding the underlying physical mechanism of the so-called super-Arrhenius viscosity and the appearance of rigidity without long range order are the fundamental challenges of glass science. 

\begin{figure*}
\centering
\includegraphics[width=1 \columnwidth, height= .4 \textheight, keepaspectratio]{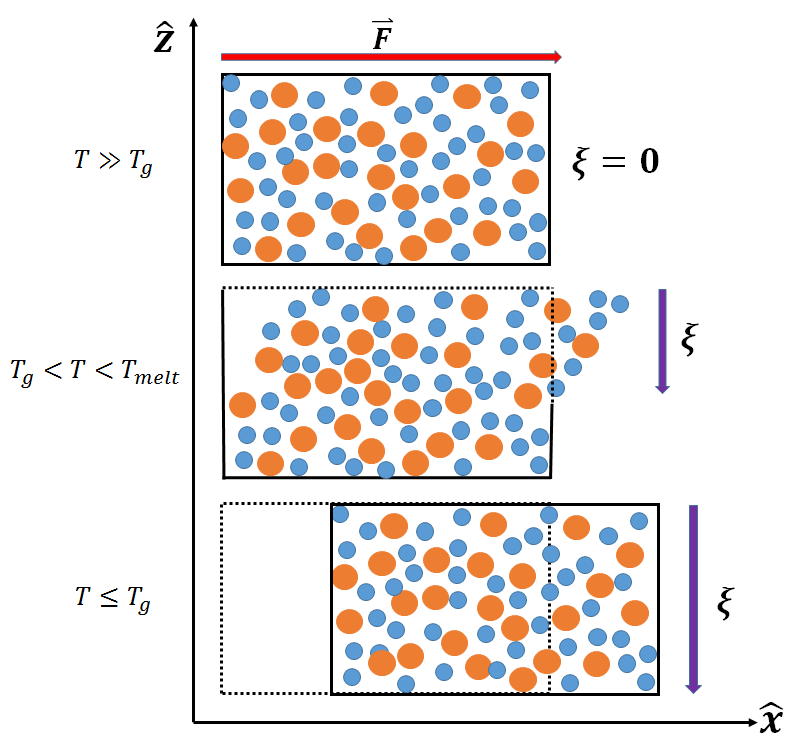}
\caption{(Color Online) Representation of the proposed response of general supercooled fluid systems. The solid lines represent the original box shape before perturbation. The dashed regions represent the successive layers that respond to the perturbation at temperatures above and around Tg. At high temperatures only the layers experiencing the external stress move appreciably, but as temperature, T, is lowered and the cooperativity becomes pronounced, the perturbation is transmitted deeper into the material, reflecting increasing rigidity. Note that the extent to which the layers move as depicted, have been greatly exaggerated for clarity.}
\label{Depth.}
\end{figure*}
An activation barrier which grows with decreasing temperature is suggestive of cooperative particle motion in the supercooled liquid, and it is natural to suspect that this cooperativity likely arises due to structural changes accompanying supercooling. Conventional wisdom posits that growing timescales require growing lengthscales \cite{9}. Further, the possible, equilibrium vanishing of the configurational entropy and associated theoretical phase transition at finite temperature, $T_K$, known as the Kauzmann paradox is highly suggestive that structre plays a significant role in the glassy slowdown. In fact, many theories of the glass transition require the existence of a growing static structural lengthscale associated with the arrest at $T_g$ \cite{1,2,3,4,5,6,10,11,12,13,14,15,16,17,18}. However, structure factors and radial distribution functions show little change upon supercooling to $T_g$, so the identification of a universally accepted structural lengthscale has remained elusive \cite{1,2,3,4,5,6}. Previously, many structural lengthscales have been proposed and tested for but unfortunately none of these lengths have been satisfactably verified experimentally \cite{19,20,21,22,23,24,25,26,27,28,29,30,31,32,33,34, Glotzer, Busch}. To that end we have proposed the existence of a new lengthscale related to structural connectivity in the supercooled liquid, namely the shear penetration depth, which is defined as the depth to which a shear perturbation applied on a boundary would penetrate into a liquid. Using molecular dynamics simulations on a model metallic glassformer, we provide evidence that this lengthscale grows dramatically with supercooling, and marks a natural candidate to measure structural connectivity in supercooled liquids tied to their arrest. The shear penetration depth has the added benefit that it can be experimentally measured.

\subsection{Facets of the Shear Penetration Depth}
We have defined the shear penetration depth as the length over which a shear perturbation applied on a boundary will appreciably penetrate into a material. If a force is applied to the boundary layer of a high temperature liquid, only the atoms in the layers of the liquid near the boundary will experience an appreciable displacement due to the applied force. As the temperature is lowered and the viscosity increases, the shear force penetrates deeper into the material until the glass transition is reached. When the material is a glass, the shear perturbation entirely penetrates the material, diverging to the system size. This idealized behavior of the shear penetration depth is depicted in Figure (\ref{Depth.}). When a crystalline solid moves together in response to a shear force, it is because the long range order allows for the transmission of the force throughout the entire material. In a glass we expect that there is also a network of interconnectivity which grew and became more cohesive as the supercooled liquid was cooled toward $T_g$. The shear penetration depth serves not only as a lengthscale for the various theories that require one, but more importantly is intimately connected with structural connectivity and its existence and growth would provide strong evidence that structure underlies the slow kinetics of the glass transition.

\section{Model and Methodology}
To test for the existence of the shear penetration depth, we simulated a model metallic glass former, namely $NiZr_2$ ($T_g \approx 700 K$). This binary alloy system represents an exceptional example of a fragile glassformer, displaying a marked super-Arrhenius viscosity. Molecular dynamics simulations were performed using the LAMMPS package \cite{48} with a cubic simulation box and periodic boundary conditions in all three cartesian directions. A velocity verlet integration algorithm with a 5 fs timestep was employed, with the force generated from a semi-empirical Finnis-Sinclair type Embedded Atom Model potential developed by Mendelev et al \cite{49}. This potential was shown to very accurately reproduce the behavior of both the equilibrium liquid and supercooled liquid/glass. 

Simulations were run with N=5000 atoms in the NPT ensemble with the target external pressure set to P=0. A Nose-Hoover thermostat and barostat were used to control the temperature and pressure, respectively. The initial positions of the atoms were generated randomly, and the system was then allowed to melt and come to equilibrium at an initial temperature of T=2200 K for 0.25 ns. Once the liquid had equilibrated, it was quenched to the desired target temperature (ranging from 300 K up to 1900 K) using a quench rate of $10^{13}$ K/s. Once the target temperature was reached, the system was allowed to evolve naturally for 0.1 ns. Next, an additional external force of 0.2 ev/A was applied in the x-direction to all atoms in the uppermost 4 angstrom layer of the simulation box. This force was left active for only 100 timesteps. This was meant to approximate an impulsive "kick" at the boundary layer of the system. After the application of the shear force, the system was left to evolve for an observation time of 16,000 timesteps. After the observation time had passed, displacement data was exported and examined. For T=300, 500, 600, 650, 700, 750, 800, and 1300 K, six independent starting configurations were employed  to provide for good statisitics. 

\section{Results and Analysis}
\begin{figure*}
\centering
\includegraphics[width=1 \columnwidth, height= .4 \textheight, keepaspectratio]{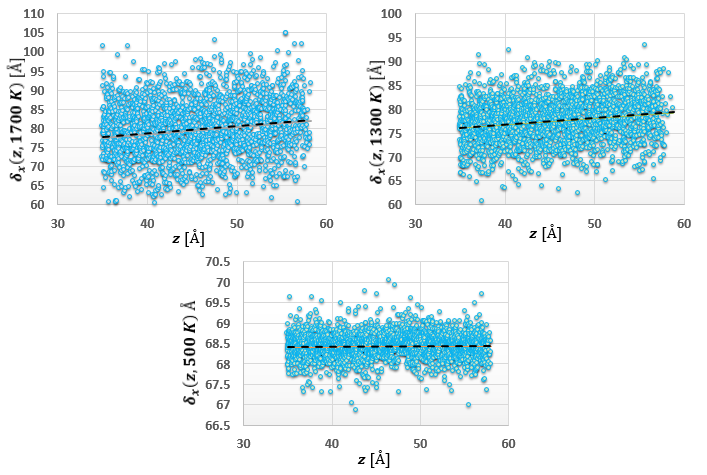}
\caption{(Color Online). Displacement data for three typical temperatures, with two above the glass transition temperature and one below.}
\label{Data.}
\end{figure*}
In order to quantify the penetration depth and extract its temperature dependence, we examine the displacement of the atoms in the shear direction as a function of their distance from the boundary layer at each of the target temperatures. Representative data is shown in Figure (\ref{Data.}). We define the shear direction (direction of applied shear) as the x-direction, and the vertical distance from the boundary layer as the z-direction (See Figure (\ref{Depth.})). The displacment is the net displacement of the atoms in the shear direction from the timestep immediately preceding the activation of the shear force to the end of the observation time, as defined above. The height of the atom is its z-coordinate at the observation time. Due to the periodic boundaries, only atoms falling within the range $z=\frac{L}{2}$ to $z=L$ are considered. Thermal and temporal effects do play a role, and make the response of the system more complicated than assumed here. This can be quantified, and a deeper analysis is saved for a separate work \cite{Me}. Because the system size is small and the applied shear force is weak (so as not to fracture the system in the glassy state), the displacement of the atoms, $\delta_x(z,T)$, as a function of height (z) can be linearized. To that end, we perform a linear regression to the data and get fits of the form,
\begin{equation}
\centering
\label{Linear}
\delta_x(z,T)=\mu *z + \delta_0.
\end{equation}
\begin{figure*}
\centering
\includegraphics[width=1 \columnwidth, height= .4 \textheight, keepaspectratio]{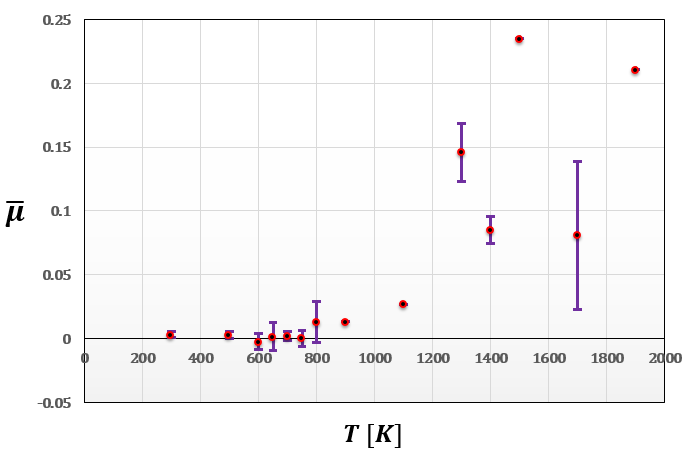}
\caption{(Color Online) Plot of the average values of the inverse penetration depth $\overline{\mu}$ as computed by Eq. (\ref{Linear}) as a function of the temperature, $T$. As seen, $\overline{\mu}$ decreases dramatically as the temperature is lowered towards the glass transition. This signifies the divergence of the shear penetration depth upon supercooling.}
\label{Slope.}
\end{figure*}
For all temperatures where multiple independent runs were made, we average the values of $\mu$, and it is this average, $\bar{\mu}$, that we examine as a function of temperature. Figure (\ref{Slope.}) depicts this temperature dependence. The error bars are given by the standard deviation computed from the multiple runs. These correspond to thermal fluctuations and the different inherent structures the supecooled liquid/glass can be quenched into.

It is clear that there is a drastic, monotonic decrease in $\bar{\mu}$ as the temperature is decreased toward $T_g$. In the vicinity of $T_g$ and at all temperatures below it, the value of $\bar{\mu}$ with its associated error is indistinguishable from zero. This is consistent with total penetration of the shear. We can convert the decay of the displacement into a quantifiable length scale by defining the penetration depth as
\begin{equation}
\centering
\label{Penetration}
\xi \equiv \frac{1}{\bar{\mu}}.
\end{equation}
Due to the fluctuations at and below $T_g$, we can only assert that the length scale grows dramatically with decreasing temperature and diverges to beyond the system size at the glass transition. A bound on the shear penetration is provided by
\begin{equation}
\centering
\label{Lower}
\xi_{l} \equiv \frac{1}{\bar{\mu}+\sigma}.
\end{equation}
\begin{figure*}
\centering
\includegraphics[width=1 \columnwidth, height= .4 \textheight, keepaspectratio]{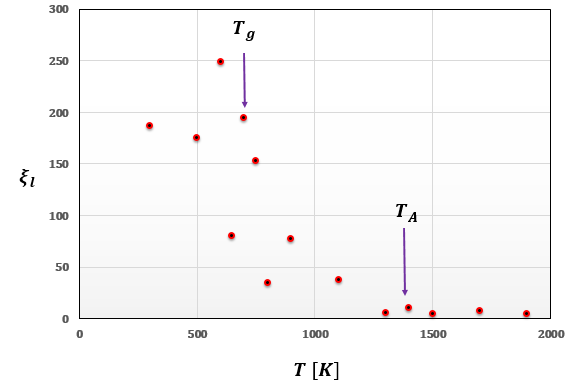}
\caption{(Color Online) The lower bound $\xi_l$ of the penetration depth (Eq. (\ref{Lower})) as a function of temperature ($T$). An increase of the shear penetration depth begins at $T_A$, see text.}
\label{Length.}
\end{figure*}
$\xi_{l}$ represents an absolute lower bound on the behavior of the length scale with temperature and is displayed in Figure (\ref{Length.}). From the figure it is clear that the length scale grows dramatically as the glass transition is approached. It is also quite noticeable that appreciable growth of the lengthscale does not set in until a temperature close to $T_A$. Below this temperature (typically for all metallic glass formers $T_A \approx 2T_g$) \cite{46} super-Arrhenius growth of the relaxation time appears. Furthermore, below $T_A$, the Stokes-Einstein relation may break down, and solid-like properties begin to appear in the liquid \cite{51,52,53,54}. More significantly, below $T_A$, local structures begin to form and percolate throughout the system as seen in numerous molecular dynamics studies \cite{52,54,63,64,65,66,67}. This provides ample support to the notion that structure plays a significant role in the arrest at the glass transition. As the local structures interlock and interpenetrate, they begin to thread the system. As they grow further, and become increasingly cohesive, a backbone forms in the liquid which is stiff and rigid. This allows for the propagation of the shear force through the material, and leads to the arrested flow. The shear penetration depth is ultimately agnostic as to the exact form of propagating structural order and connectivity, but represents an exciting new method of possibly measuring this growing order without resorting to standard scattering methods which have been largely opaque to role of structure in the glass transition.

\section{Conclusion}
By taking advantage of simple physical reasoning derived from the main distinction between liquids and solids, in terms of their response to shear, we have identified a natural length scale relevant to the glass transition. This lengthscale, the shear penetration depth, has been shown through simulations, to grow dramatically as a liquid is supercooled toward $T_g$. Further, we have made a reasonable and physically meaningful connection with growing structural order and connectivity in the supercooled liquid. That the shear penetration depth begins to grow appreciably at $T_A$, is a strong indicator that structure plays a fundamental role in the kinetic phenomenology of the glass transition. Because the shear penetration depth is intimately related and sensitive to this connectivity and structural change, it marks an exceptional candidate for revealing the effect of structure in experiments. 
\section{Acknowledgements}
NW and ZN were supported by the NSF DMR-1411229. ZN thanks the Feinberg foundation visiting faculty program at Weizmann Institute.

\end{document}